\documentclass[proceedings]{JHEP3}

\PrHEP{PrHEP unesp2002}                   
\conference{Workshop on Integrable Theories, Solitons and Duality}                

\def\l{\lambda}
\def\be{\begin{equation}}
\def\ee{\end{equation}}
\def\bea{\begin{eqnarray}}
\def\eea{\end{eqnarray}}

\usepackage{epsfig}

\title{Noncommutative Field Theories and Integrable Models in 2d}

\author{\speaker{Marco Moriconi}\\                      
Newman Laboratory of Nuclear Studies, Cornell University\\
Ithaca, New York 14853, USA\\ 
and\\ 
Instituto de F\'\i sica\\
Universidade Federal do Rio de Janeiro\\
Rio de Janeiro, RJ 21945-970, Brazil\\
E-mail: \email{marco@if.ufrj.br}}                 
\author{I. Cabrera-Carnero\\                           
IFT, Unesp\\
Rua Pamplona, 145\\
S\~ao Paulo, SP 01405-900, Brazil\\   
E-mail: \email{cabrera@ift.unesp.br}}

\abstract{We study the noncommutative extensions of certain integrable field theories, namely the 
sine- and sinh-Gordon (sG and shG) models, and the $U(N)$ principal chiral model (pcm). We argue that the Moyal deformations of the sG and shG models are not integrable, by looking at tree-level amplitudes where there is particle production. By considering the noncommutative generalization of the zero-curvature method, it is possible to define integrable versions of the noncommutative sG and shG models, which introduce extra constraints. The noncommutative pcm is shown to be integrable and we discuss the existence of non-trivial non-local conserved charges, and the associated noncommutative zero-curvature condition.}

\begin{document}

  \section{Introduction}

In this note we are going to summarize the results described in \cite{paper}. The presentation
will be rather informal and we refer the interested reader to that reference in order to
see a more complete discussion.

Noncommutative field theories (ncfts) have attracted a great deal of attention in the past few years
due to several reasons. First and foremost, its connection to string theory, where they arise
as certain limits of type IIB theories with a B field turned on \cite{SW},
they serve as toy-models for the study of quantum gravity since they are intrinsically nonlocal, and a connection between noncommutative Chern-Simons theory and the fractional quantum Hall effect has been suggested by Susskind \cite{Suss}. Finally, there is a host of theoretical phenomena which are rather different from usual quantum field theory (qft), making them very
appealing from the theoretical point of view \cite{MvRS}.

We will be considering euclidean field theories, since it has been pointed out that the introduction of
space and time noncommutativity may lead to violations of unitarity \cite{GM}. It is conceivable tough, that some specific models could evade these arguments \cite{CLZ,BDFP}.

\section{Generalities on Noncommutative Field Theories}

We can define a noncommutative deformation of a given model, defined by some local
lagrangian, by replacing the products of fields by the Moyal product, also known as the $\star$-product,
\begin{equation}
 \left. \phi(x)\eta(x) \to \phi(x) \star \eta(x) = \exp(\frac{i}{2}\theta^{\mu \nu}\partial^{z_1}_\mu\partial^{z_2}_\nu)
\phi(z_1)\eta(z_2)\right|_{z_1=z_2=x}
\end{equation}
Once this has been done, we can deform the whole action and from there, derive the new Feynman rules \cite{Filk,reviews}. The first important point to notice is that quadratic terms in the action are equivalent to non-deformed terms, provided the fields fall fast enough at infinity and that we work in manifolds without boundaries and with trivial topology. This is due to the fact that $\phi(x)\star\eta(x)=
\phi(x)\eta(x)+\partial_\mu T^\mu(x)$ for some (nonlocal) operator $T^\mu(x)$. Therefore the propagator of a ncft is the same as the non-deformed version ($\theta=0$). The only difference will be in the vertices of the theory. One point that should be made though, is that in most cases the measure of the path integral is taken to be the same for $\theta \neq 0$ as the one for $\theta=0$, which amounts to saying that the vacuum of the deformed theory is the same as the one of the non-deformed theory. This can be established perturbatively, but one should be aware of possible complications due to non-perturbative effects.

We are going to use a $\star$ subscript to indicate that we are considering Moyal products. For example 
$\phi^4_\star=\phi \star \phi \star \phi \star \phi$.

Let us analyze the change in the case of the $\phi^4$ and $\phi^6$ vertices.
In momentum space the space-time integral of the Moyal product of a string of fields becomes
\begin{equation}
\int dx \phi^n_\star=\int dp_1 dp_2 \ldots dp_n \exp(-\frac{i}{2}\sum_{i<j}(p_i)_\mu\theta^{\mu \nu}(p_j)_\nu) \phi(p_1)\phi(p_2)\ldots\phi(p_n) \delta(p_1+p_2+\ldots+p_n)
\end{equation}
We will introduce now the following convention: $\frac{1}{2}(p_i)_\mu \theta^{\mu \nu} (p_j)_\nu \equiv p_i \wedge p_j)$ \footnote{Note a difference of a factor of $\frac{1}{2}$ from the usual convention in the literature.}.
In the case of scalar $\phi^4$ vertex, this expression can be simplified by symmetrizing it over the momenta
\bea
&&\int dx \phi^4_\star = 
\frac{1}{4!}\sum_{perm.} \exp(-i\sum_{i<j} p_i \wedge p_j)=
\frac{1}{3}(\cos(p_1\wedge p_2) \cos(p_3 \wedge p_4) + \nonumber \\
&&\phantom{\int dx \phi^6_\star =}
\cos(p_1\wedge p_3) \cos(p_2 \wedge p_4)+\cos(p_1\wedge p_4) \cos(p_2 \wedge p_3)). \label{v4}
\eea
The $\phi^6$ vertex becomes
\be
\int dx \phi^6_\star=
\frac{1}{6!}\sum_{perm.}\exp(-i\sum_{i<j}p_i \wedge p_j). \label{v6}
\ee
Therefore all one needs to do in order to compute anything {\em perturbatively} is to replace the vertices of a given Feynman
diagram by their noncommutative expressions, like \ref{v4} and \ref{v6}, and use the same propagators as in the commutative theory.

\section{Integrable Field Theory}

An integrable field theory is characterized by an infinite number of conserved charges in involution. This imposes severe constraints in the theory: there is no particle production in scattering processes, the set of in- and out-momenta is the same, the amplitude for $n$-particle scattering processes can be factorized in terms of two-body processes, introducing the all important two-body $S$-matrix $S_2$. The two-body $S$-matrix satisfies the Yang-Baxter equation (YBe) as a consequence of integrability. The YBe is trivial for theories where $S_2$ is a collection of phases, but it is highly non-trivial for theories where $S_2$ has a non-diagonal structure.
This makes integrable field theories more controlled technically, without making them trivial \cite{ZZ,D}. Aside from their theoretical appeal, they have found applications in a great number of different problems, in statistical mechanics, condensed matter, and high-energy physics, via string theory.

Out of the consequences of integrable field theories, we can say that the hallmark of integrability is the non-production of particles in any scattering process. This has to be true to all orders in perturbation theory, which in a loop expansion
means that the tree-level amplitudes for particle production must vanish. This encodes the fact that the classical theory is integrable.

The models we are going to study are: sine-Gordon (sG), sinh-Gordon (shG), and principal chiral model (pcm). The sG model is defined by the following lagrangian
\be
{\cal L}_{sG}=\frac{1}{2}(\partial \phi)^2+\frac{m^2}{\beta^2}(\cos(\beta \phi)-1) \ . \ \label{sG}
\ee
The shG model is obtained from the sG model by replacing $\beta \to i\beta$,
\be
{\cal L}_{shG}=\frac{1}{2}(\partial \phi)^2+\frac{m^2}{\beta^2}(1-\cosh(\beta \phi)) \ . \ \label{shG}
\ee
Finally, the pcm is defined through the following lagrangian
\be
{\cal L}_{pcm}=\frac{1}{2g_0^2}\rm{Tr}(\partial^\mu g^{-1}\partial_\mu g) \ , \ \label{pcm}
\ee
where $g$ is a group element, of one of the classical groups. Here we are going to consider only the $U(N)$ group,
due to reasons that will become clear later. We only stress the fact that noncommutativity imposes constraints that restrict the existence of certain groups \cite{A}.

The sG and shG models are related by a simple replacement in the coupling constant, but have very different vacuum structure. In the shG model there is a single minimum, whereas in the sG model there is an infinite degeneracy, implying the existence of solitons. In any case, since the scattering amplitudes are analytical in the coupling constant, we can restrict our attention to the shG model, for simplicity (see \cite{NOS} for a discussion of the noncommutative sine-Gordon model).

We will consider a simple scattering process of $2$ particles going into $4$ in the shG model at tree-level, whose amplitude is denoted by ${\cal M}_{2 \to 4}$. For this computation we can truncate the lagrangian \ref{shG} at the $\phi^6$ term
\be
{\cal L}_{shG}^{(6)}=\frac{1}{2}(\partial \phi)^2-\frac{m^2}{2}\phi^2-\frac{m^2\beta^2}{4!}\phi^4-\frac{m^2\beta^4}{6!}\phi^6
\ee
There are three classes of diagrams in the computation of ${\cal M}_{2 \to 4}$, which are depicted in figure 1.
In performing this computation we have to symmetrize the sums over the final momenta and take energy-momentum constraints into consideration. For $\theta = 0$ these graphs add up to 0. This is due to the fact that the shG model is integrable classically. In the deformed theory there is a non-zero contribution at order $\theta^2$, which means that the amplitude for particle production is non-zero. This means that the na\"\i ve Moyal deformation of the shG (and sG) model is {\em non}-integrable.
\EPSFIGURE{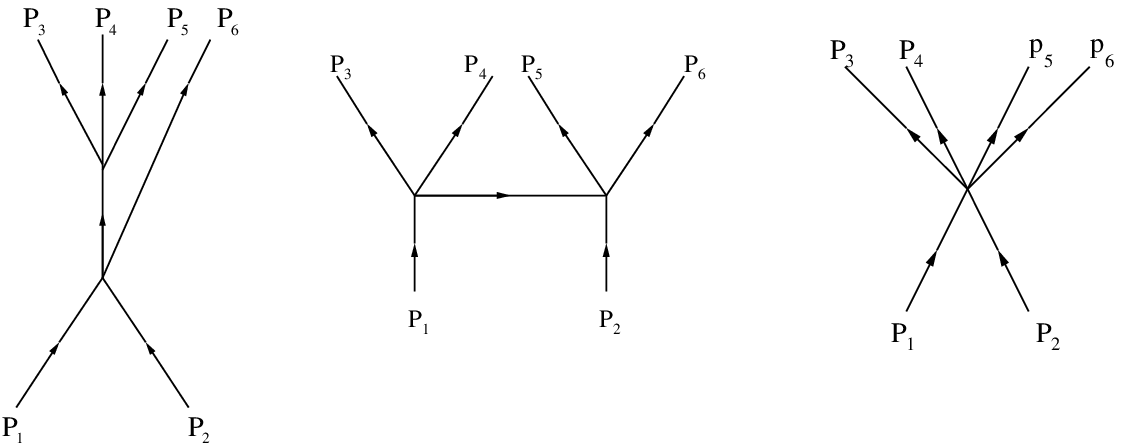}                       
{The three classes of diagrams contributing to ${\cal M}_{2 \to 4}$.\label{fig}}
This way of defining the noncommutative sG and shG models gives 
the following equation of motion for the noncommutative sG
\be
\partial^\mu\partial_\mu\phi+\frac{m^2}{\beta}\sin_\star(\beta\phi)=0
\ee
with no further constraints. 

On the other hand we can define these models through a different route such that the following two requirements are satisfied: 1- as $\theta \to 0$ we obtain the usual sG model, and 2- there is an infinite number of conserved charges by construction. We will see that this is possible to be done by generalizing the zero-curvature method to noncommutative spaces.

\section{Noncommutative Zero-Curvature Method}

If one manages to cast the equations of motion of a given two-dimensional model in the following form
\be
\partial_t U-\partial_x V + [U,V]=0 \label{zc}
\ee
where $U$ and $V$ are matrices depending on the fields and their derivatives, and a spectral parameter $\lambda$, and $[U,V]=UV-VU$,
then it is possible to show that there is an infinite number of conserved charges in this model \cite{FT}. When writing the equations of motion in this form, in general the diagonal terms will be the equations of motion and the off-diagonal terms vanish. We can generalize the zero-curvature condition \ref{zc} by changing the commutator of $U$ and $V$ by their $\star$-commutator
\be
\partial_t U-\partial_x V + [U,V]_\star=0 \label{zcs}
\ee
where $[U,V]_\star \equiv U\star V-V\star U$. Once the equations of motion have been cast in this form, we can follow the derivation of the existence of an infinite number of conserved charges in the commuting case line by line \cite{paper}. The upshot of this discussion is that the operator $T_\l(x)$, defined as the solution of
\be
\frac{\partial T_\l}{\partial x} = U \star T_\l \label{t}
\ee
can be used to generate an infinite number of conserved charges. Namely, we can show from the noncommutative zero-curvature condition that $\partial_t{\rm Tr}T_\l(L)=0$, and by expanding $T_\l(L)$ in powers of the spectral parameter, we obtain the sought for conserved charges.

We will work in light-cone coordinates now, where $x_\pm=(x_0 \pm x_1)/2$.  
It is possible to write the equation of motion for the shG model as a zero-condition
equation, by introducing $A$ and $\bar A$,
\be
A=-\frac{m\lambda}{2}(e^{\beta\phi} \,\, \sigma_- 
+e^{-\beta\phi} \,\, \sigma_+)
\qquad {\rm and} \qquad
{\bar A}=\frac{m}{2\lambda}(\sigma_-+\sigma_+)-
\frac{\beta}{2}\bar\partial \phi \,\, \sigma_3 \label{aabar}
\ee
where $\sigma_{\pm}=\frac{1}{2}(\sigma_1\pm i\sigma_2)$, $\sigma_i$
are the usual Pauli matrices, and $\lambda$ is the spectral parameter.
It is a simple computation to show that the zero-curvature condition for $A$ and $\bar A$
\be
\bar \partial A - \partial \bar A + [A,\bar A]=0 \ , \  \label{zc_lc}
\ee
is equivalent to the equation of motion for the shG model. Notice that in
showing this, the diagonal elements of the matrix equation
are the equation of motion and the off-diagonal elements vanish.
 
We {\em define} the noncommutative shG model through the
$\star$-zero-curvature equation
\be
\bar \partial A- \partial \bar A + [A,\bar A]_{\star}=0 
\ , \ \label{AA}
\ee
where $[A,\bar A]_{\star}=A\star\bar A-\bar A\star A$.  The equation
of motion derived from \ref{AA} is
\be
\partial\bar\partial\phi+\frac{m^2}{\beta}\sinh_\star(\beta\phi)=0 \label{eq1}
\ee
which is exactly the same equation one would obtain from the Moyal deformation
of the shG action. There are, though, two more constraints, coming from the
off-diagonal elements, and which read
\bea
&&\bar \partial (e^{-\beta \phi}_\star)+
\frac{\beta}{2}(e^{-\beta\phi}_\star \star \bar \partial \phi +
                \bar\partial\phi \star e^{-\beta\phi}_\star)=0 \\
&&\bar \partial (e^{\beta \phi}_\star)-
\frac{\beta}{2}(e^{\beta \phi}_\star \star \bar \partial \phi +
\bar\partial \phi \star e^{\beta\phi}_\star)=0  \ . \ \label{constraints}
\eea
It is easy to show that these constraints can be written as total derivatives. Therefore the extra constraints can be seen as global conserved currents. They clearly vanish as $\theta \to 0$,
when the Moyal product becomes the usual product, and we obtain the equations of motion of the shG model, as expected.

In \cite{GP} those authors proposed a definition of the noncommutative sG model using the bicomplex method. In spirit this is similar to what we have done, since by construction they generate an infinite number of conserved charges, and as $\theta \to 0$ they recover the equation of motion of the sG model. They also obtained two extra constraints which can be written as total derivatives (global currents). The equation of motion found in \cite{GP}, after replacing $\beta$ by $i\beta$ is
\be
\partial(e^{\beta\phi}_\star \star \bar\partial(e^{-\beta\phi}_\star)-
         e^{-\beta\phi}_\star \star \bar\partial(e^{\beta\phi}_\star))=
2 m^2\sinh_\star(\beta \phi)
\ee
which is different from the one we found here. It is possible, though, to make a different choice of potentials for $A$ and $\bar{A}$ such that we obtain the same equation of motion. This is due to the fact that when we generalized these gauge potentials to the noncommutative case, we kept the $\partial \phi$ term intact, but this choice is not unique: we could have chosen $\frac{1}{2\kappa}(e_\star^{-\kappa\phi} \star \partial e_\star^{\kappa\phi}- e_\star^{\kappa\phi} \star \partial e_\star^{-\kappa\phi})$. As $\theta \to 0$ the product of fields becomes the usual product and we trivially recover $\partial \phi$. If we take $\kappa=1$, the zero-curvature condition with the gauge potentials becomes precisely the equation of motion proposed by Grisaru and Penati. It is harder to check that the same holds for the constraints. It would be very interesting to establish whether they are the same or not.
 
\section{Euclidean Solitons}

Now that we have the equations of motion for the noncommutative sG model we can try to solve them and see what is the generalization of the solitons in the original theory. The somewhat surprising result is that the $1$-soliton solution of the sG model {\em solves} the equation of motion and constraints of the noncommutative sG model. This is due to the fact that the $\star$-product of two functions that depend linearly on the space-time variables is the same as their usual product. There are several ways to prove this, for example, consider the $\star$-product of $f(\sum_\alpha a_\alpha x^\alpha)$ and $g(\sum_\alpha a_\alpha x^\alpha)$
\be
\left.f(\sum_\alpha a_\alpha x^\alpha) \star g(\sum_\alpha a_\alpha x^\alpha)=\exp(\frac{i}{2}\theta^{\mu \nu}\partial_\mu^{z_1}
\partial_\nu^{z_2})f(\sum_\alpha a_\alpha z_1^\alpha)g(\sum_\alpha a_\alpha z_2^\alpha)\right|_{z_1=z_2=x}
\ee
Expanding the exponential, using the chain rule, and the antisymmetry of $\theta^{\mu \nu}$ we see that all terms but the first vanish. This is true in any dimensions, but we are interested in the two-dimensional case only. Therefore the equation of motion of the noncommutative sG model is the same as the one of the usual sG model. Moreover, from the form of the constraints, replacing the $\star$-products by their usual product, implies the vanishing of these constraints. All this put together means that if a certain field solution depends linearly on the spacetime variables, which is the case for soliton solutions in two-dimensions, then it will also solve the noncommutative equations of motion. This establishes the fact that the $1$-soliton solution of the sG mode is also a solution of the noncommutative sG model.

\section{Noncommutative Principal Chiral Model}

We have seen an example where the na\"\i ve Moyal deformation of an integrable field theory does not provide an integrable field theory. We will see now an example where the Moyal deformation of the action of an integrable field theory provides one, the principal chiral model (pcm) \cite{P}.

The action of the pcm is
\be
{\cal S}_{pcm}=\frac{1}{2 g_0^2}\int dx {\rm Tr} (\partial^\mu g \partial_\mu g^{-1})
\ee
where $g$ is a groups element, belonging for example to $U(N)$ or $SU(N)$, and satisfies $gg^\dagger=g^\dagger g=1$. 
The Moyal deformation of the pcm is given by the action
\be
S^*_{pcm}=\frac{1}{2g^2_0}\int d^2x {\rm Tr}
(\partial^{\mu}g^{-1} \star \partial_{\mu} g) \label{pcm*}
\ee
and the field $g$ is required to satisfy $g \star g^{\dagger}= g^{\dagger}\star
g =1$. The reason why we should be specific about the group to which $g$
belongs to, is that not all groups allow noncommutative extensions, for example,
there is no noncommutative $SU(N)$. Therefore we will restrict our analysis
to the $U(N)$ pcm. Since the action \ref{pcm*} is quadratic and the manifold we are considering is trivial, the actions \ref{pcm} and \ref{pcm*} are identical \footnote{Provided the fields fall fast enough at infinity.}. The only difference between these two models lies in the constraints. We can construct non-trivial non-local conserved charges by following the 
Brezin-Itzykson-Zinn-Justin-Zuber (BIZZ) method \cite{BIZZ}, and we refer to\cite{paper} for this discussion.

The noncommutative zero-curvature method can be applied to the noncommutative pcm, but this time there are no extra constraints. Consider the potentials
\bea
&&U(\lambda)=\frac{1}{2}\frac{l_0+l_1}{1-\lambda}-
\frac{1}{2}\frac{l_0-l_1}{1+\lambda} \\
&&V(\lambda)=\frac{1}{2}\frac{l_0+l_1}{1-\lambda}+
\frac{1}{2}\frac{l_0-l_1}{1+\lambda} \ , \
\eea
where
\be
l_0(x,t)=\frac{\partial g}{\partial t}*g^{-1}
\qquad {\rm and} \qquad
l_1(x,t)=\frac{\partial g}{\partial x}*g^{-1} \ . \
\ee
Introducing this on the zero-curvature condition \ref{zc} we obtain
\be
\frac{\partial^2 g}{\partial t^2}-\frac{\partial^2 g}{\partial x^2}=
\frac{\partial g}{\partial t}*g^{-1}*\frac{\partial g}{\partial t}-
\frac{\partial g}{\partial x}*g^{-1}*\frac{\partial g}{\partial x}
\ee
which is the equation of motion of the pcm, and it can be rewritten in
the more compact form
\be
\partial_{\mu}(g^{-1}*\partial^{\mu}g)=0 \ . \
\ee
Contrary to the noncommutative sG model, there are no further constraints
in the noncommutative $U(N)$ pcm.

\section{Conclusions}

We have seen that the Moyal deformation of a given 2d integrable model does 
not necessarily provide a integrable field theory. In the case of the 
sinh-Gordon model (and by replacing $\beta \to i\beta$, the sine-Gordon model) 
we were able to establish their {\em non}-integrability by
computing the amplitude for $2 \to 4$ particles at the tree-level, and verifying it is non-zero. 
On the other hand, the noncommutative $U(N)$ principal chiral model defined through the Moyal
deformation of the action and constraints of the $U(N)$ principal chiral model, does provide an
integrable field theory, where the elegant method of Brezin et al \cite{BIZZ}
works as well as in the commutative case.

The equations of motion we found for the noncommutative sG model initially are different from the
ones proposed by Grisaru and Penati in \cite{GP}. Upon a change in the definition of the noncommutative
version of $\partial \phi$ we were able to find the same equation of motion as \cite{GP}, 
but it is not trivial to establish the equality of the constraints.

We showed that the 1-soliton solution of the sine-Gordon model solves the equations of motion and constraints  of the noncommutative version. It would be interesting to study the more general multi-soliton solutions.

There are several interesting directions to pursue. Initially, it would be nice to have a more thorough
understanding of the conservation laws, verifying for example, that these charges are in involution. Next one could consider the noncommutative versions of different models, such as the affine Toda theories. And last but not least, the investigation of the quantization of these models is a fascinating, if somewhat difficult problem.

\section{Acknowledgments}
We would like to thank the hospitality of the Abdus Salam ICTP, where
we started this work. Discussions with P. Dorey, J. Evans, A. Garcia,
K. Narayan, A. Leclair, M. Neubert, V. Sahakian, M.M. Sheikh-Jabbari, and
J. Varilly are gracefully acknowledged. One of us (MM) would like to thank 
F. Muller-Hoisen for email exchanges, and  C. Wotzasek for the use
of computer facilities. We also would like to thank the organizers of this workshop for the hospitality and to  wish the IFT and its members many happy returns to this date. 
This work is in part supported
by the NSF and CNPq (profix) (MM) and Fapesp (ICC).



\end{document}